\documentclass[APA,STIX1COL]{WileyNJD-v2}
\usepackage{multicol}
\usepackage{graphicx}
\usepackage{float}

\articletype{Article Type}%

\received{1 August 2024}
\revised{1 August 2024}
\accepted{1 August 2024}

\def\printjnlcitation{}

\begin{document}

\title{Revolutionizing Bridge Operation and Maintenance with LLM-based Agents:
 An Overview of Applications and Insights}

\author[1]{Xin-yu Chen}

\author[1,2,3]{Lian-zhen Zhang*}

\authormark{AUTHOR ONE \textsc{et al}}

\address[1]{\orgdiv{School of Transportation Science and Engineering}, 
\orgname{Harbin Institute of Technology}, \orgaddress{\state{Harbin}, \country{China}}}

\address[2]{\orgdiv{School of Transportation and Civil Engineering}, 
\orgname{Shenzhen University}, \orgaddress{\state{Shenzhen}, \country{China}}}

\address[3]{\orgdiv{National Key Laboratory of Green Longevity Road Engineering for Extreme Environments}, 
\orgname{Ministry of Science and Technology}, \orgaddress{\state{Shenzhen}, \country{China}}}

\corres{*Lian-zhen Zhang, School of Transportation Science and Engineering,
Harbin Institute of Technology,Harbin510001,China. \email{lianzhen@hit.edu.cn}}

\presentaddress{Harbin Institute of Technology,Harbin510001,China}

\abstract[Summary]{In various industrial fields of human social development, people are constantly 
exploring technologies to free human labor. The construction of LLM-based agents is considered to be 
one of the most effective tools to achieve this goal. LLM-based agents, as a kind of human-like 
intelligent entities with the ability of perception, planning, decision-making, and action, have 
already created great production value in the fields. As the most important transportation 
infrastructure, how to keep bridges in safe service is a major industry need, and research on 
intelligent operation and maintenance technologies is urgently needed. In general, the bridge 
operation and maintenance field show a relatively low level of intelligence compared to other 
industries. Nevertheless, numerous intelligent inspection devices, machine learning algorithms, 
and autonomous evaluation and decision-making methods have been developed in the field of bridge 
operation and maintenance, which provides a feasible basis for breakthroughs in this field. This 
study aims to explore the impact of LLM-based agents on the field of bridge operation and maintenance 
and to analyze the potential challenges and opportunities they bring to the core tasks. Through 
in-depth research and analysis, this paper expects to provide a more comprehensive perspective for 
understanding the application of LLM-based agents in the field of bridge intelligent operation and 
maintenance.
}

\jnlcitation{\cname{%
\author{X.-Y. Chen},
\author{Y.-W Zhu},
\author{Y. Hou},
\author{L.-Z. Zhang}} (\cyear{2024}),
\ctitle{Revolutionizing Bridge Operation and Maintenance with LLM-based Agents:
An Overview of Applications and Insights}, \cjournal{:}, \cvol{2024;}.}

\maketitle

\footnotetext{\textbf{Abbreviations:} LLMs, Large Language Models; O\&M, Operation and Maintenance}

\setlength{\columnsep}{20pt}
\begin{multicols}{2} 
\section{Introduction}\label{sec1}
Within the past decade or so, China has made great achievements in the construction of highway 
bridges, with the maximum span breaking through and the number of bridges increasing. By the end 
of 2023, there were 1,079,300 existing highway bridges in China. At the same time, a large number 
of bridge structures have entered the middle and late stages of service one after another, and safety 
accidents are frequent, causing widespread concern in society. Optimizing the operation and 
maintenance of existing bridges to maintain and improve their safety performance and service 
status has become an urgent need for the society.

Highway bridge operation and maintenance enhance structural safety and efficiency 
through technological innovation and management strategies, realizing sensor monitoring
~\citep{he2022integrated,rizzo2021challenges}, drone 
detection, algorithmic optimization of disease identification, scientific evaluation methods
~\citep{sasmal2008condition,catbas2002condition}, and 
multilevel maintenance decision-making system
~\citep{han2021risk,bocchini2011probabilistic}, which lays the foundation for the intelligent 
development of the industry. However, there are many problems, such as poor stability of bridge 
perception data and high data processing delay. The technical evaluation dimension is single, and 
the model accuracy is low. Operation and maintenance decision-making lack of data support, low level 
of intelligence. Bridge operation and maintenance intelligence is still in the primary stage, need 
to strengthen the state perception, evaluation system, decision support and integration with 
artificial intelligence technology.

The problems faced by bridge O\&M urgently require LLM-based agents technology to solve them.
Agents, as human-like intelligent entities, are capable of perception, planning, decision-making 
and action autonomously~\citep{woodridge1995intelligent,goodwin1995formalizing}. 
Although limited to specific tasks such as Go games or book retrieval in 
the early days, the Transformer architecture introduced by \cite{vaswani2017attention} has greatly boosted the 
development of Language Models. Large Language Models such as OpenAI's ChatGPT and Google Bert's 
have brought new hope to the study of agents by significantly improving comprehension, summarization, 
reasoning, and language processing with their excellent generalization capabilities. These Large 
Language Models are now regarded as the “brain” of agents, and have great potential for application 
in the field of bridge operation and maintenance.

In the context of the current rapid development of the research field of Large Language Models and 
Agents, the emergence rate of new Large Models and Agents has increased significantly. The academic 
community has paid great attention to this, and the publication of related research results has shown 
a sharp increase. This shows that AI technology is gradually penetrating and transforming various 
industries. Therefore, it is necessary to explore and introduce the corresponding advanced technology 
in the field of bridge operation and maintenance in order to promote the innovative development of 
this field.

At the current research stage, agents are more mature in perception, planning, decision-making, and 
action. However, the bridge O\&M domain is not sufficiently automated and intelligent in terms of 
disposition sensing, data prediction, performance evaluation, decision constraints, emergency 
response, and disaster mitigation. Therefore, the new generation bridge O\&M management system 
should integrate LLM-based agent technology to equip each bridge with a personal assistant. At 
the same time, the system should consider the complementary and independent relationships of 
bridge clusters at the road network level to realize synergy, cooperation and competition among 
agents. The purpose of this paper is to discuss how  agents can optimize bridge O\&M tasks, how to 
build a bridge O\&M intelligent system, and to analyze the impact of AI on the future development of 
the bridge O\&M field, as well as the opportunities and challenges that may be encountered with the 
development of agents.

\begin{center}
    \includegraphics[width=0.4\textwidth]{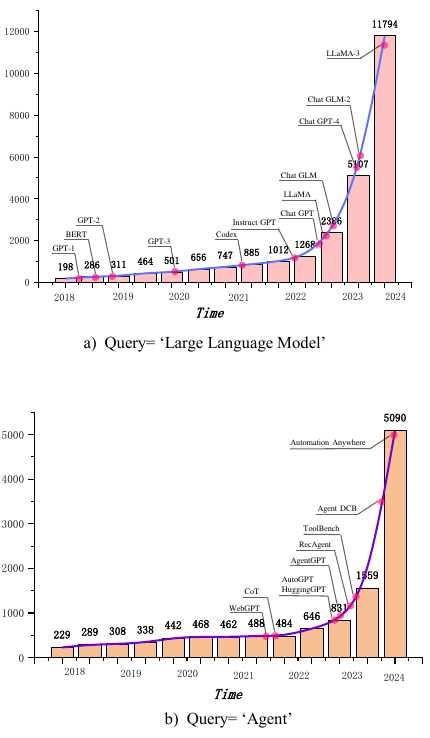}
    \makeatletter
    \def\@captype{figure}
    \makeatother
    \caption{Trends in the number of arXiv papers published containing the 
    keywords "Large Language Model" and "Agents" title or abstract by month, 
    using exact matches. We set different x-axis ranges for the two keywords. 
    We labeled the points corresponding to important milestones in the progress of LLM and 
    Agents' research.The number of papers containing "Large Language Model" in the title or 
    abstract increased dramatically after the release of ChatGPT (Fig.a). 
    The number of papers containing "A" in the title or abstract increased dramatically after 
    the release of AuotoGPT and HuggingGPT increased (Fig.b).\label{fig1}}
\end{center}

\section{Background}\label{sec2}
In this section, we provide a detailed overview of the evolution of bridge operation and maintenance and LLM-based agents 
research. We find that the shortcomings of modern bridge health monitoring techniques match the 
advantages of LLM-based agents. By analyzing the development trajectories of both, we propose the 
use of LLM-based agents to invigorate the field of bridge operation and maintenance and to promote industrial innovation 
in the current productivity context.

\begin{figure*}[!t]
    \centerline{\includegraphics[width=34pc,height=21pc]{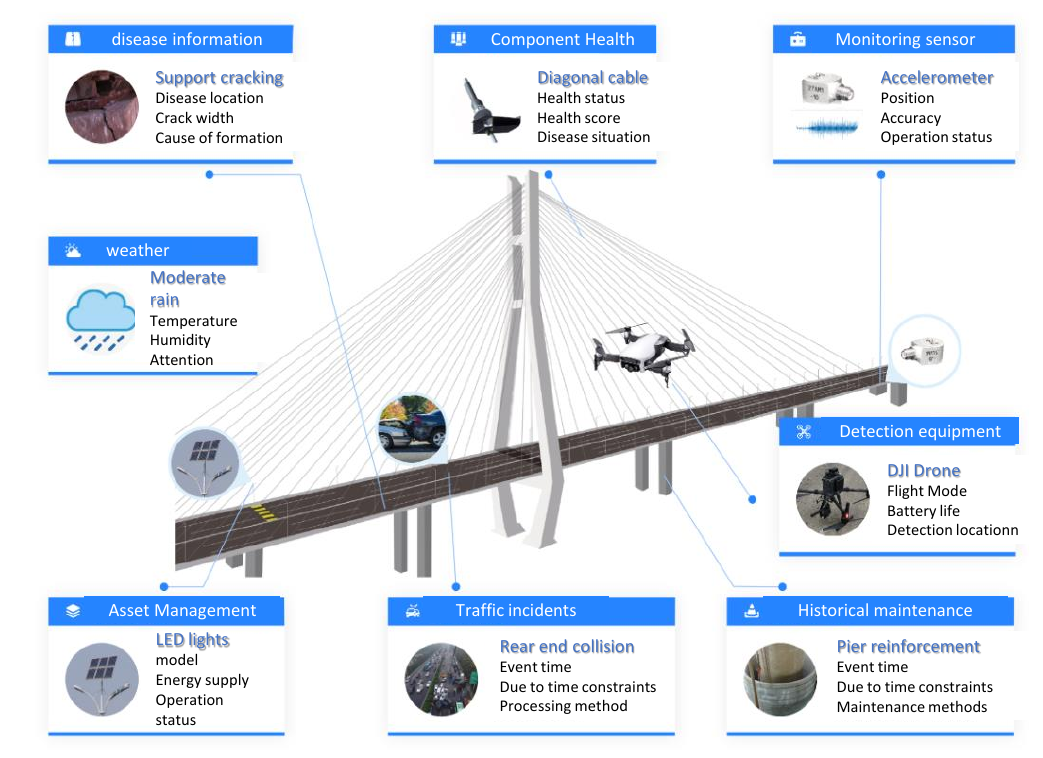}}
    \caption{Modern bridge digital inspection and monitoring technology. 
    It allows observation of the environment in which the bridge is located, the loads on the 
    structure, the changes in the structure, and the load response of the structure. 
    \label{fig2}}
\end{figure*}

\subsection{Review of Bridge Operation and maintenance Research}

The Bridge Management and Maintenance System is responsible for the health management of bridges 
throughout their life cycle, with functions including data collection, integration and storage, 
equipment management, condition assessment, performance prediction, strategy development and 
emergency response. The system is planned during the construction phase to monitor critical areas 
and sensitive parameters. After the bridge is completed, the difference between the actual operating 
condition of the bridge and the design expectation is assessed based on factors such as material 
aging, structural characteristics and traffic load. It is worth emphasizing that the goal of a 
bridge management and maintenance system is not to keep bridges in optimal condition at all times, 
but rather to ensure that the long-term operation of bridges produces the maximum economic benefit 
to society, within the constraints of limited financial budgets, bridge substitutability, road 
network importance and accident prevention.

After the occurrence of major accidents on bridges, people began to realize that bridges need to 
establish a special system, which is responsible for the regular inspection and periodic maintenance 
of bridges to ensure the normal use of bridges and to reduce the occurrence of vicious events. 
The development of bridge health monitoring systems is divided into three main stages:

The first phase was from the 1960s to the 1980s, when bridge maintenance records were based on paper 
documents. During this period, bridge information was updated infrequently, and documents were poorly 
circulated and standardized. Typically, bridge maintenance activities were carried out only when 
there was significant damage or accidents to the structure. During this period, bridge maintenance 
systems were mainly found in countries where bridge construction was carried out early, such as 
Sweden and the United States~\citep{soderqvist1998finnish}.

The second stage is from the 1980s to the end of the twentieth century, when the bridge maintenance 
system was constructed as a more complete software system. This stage is marked by the improvement 
of the bridge maintenance software system, the development of which was born out of the original 
paper-based document system, and evolved with the progress of computer technology. Maintenance 
information is gradually transferred from the traditional paper media to computer storage, the 
main forms include text, tables and images. Software system functions are also gradually enriched, 
covering bridge information management, daily monitoring and maintenance records, and maintenance 
decision support and other aspects. The most representative system at this stage is the U.S. PONTIS
~\citep{thompson1998pontis}, 
after the completion of the system in the United States, people realize the importance of building 
bridge maintenance software, each country began to build their own bridge maintenance system, the 
representatives of Europe, including Denmark's Danbro, the United Kingdom's NATS, France's Edouard, 
Norway's Brutus~\citep{mandic2019european}, 
the representatives of Asia, including the Japan's J- BMS of Japan, BMS of Korea 
and CMBS of China~\citep{jeong2018bridge}.

The third stage is from the twenty-first century to the present, where the effective integration 
of novel algorithms and theories has realized the double improvement of bridge structural safety 
and operation and maintenance efficiency.

There are mainly three kinds of innovations in algorithms and theories. (1) In the development 
of algorithms, the integration of high-precision image processing, laser point cloud 3D 
reconstruction and holographic photography technology has improved the accuracy of the 
identification of bridge surface diseases and real-time detection of structural service state; 
(2) In the evaluation of bridge technical condition, scholars have developed a variety of evaluation 
methods, including the fuzzy theory, hierarchical analysis and disease weighting system, in order 
to scientifically assess the technical condition of bridges; (3) In the research of maintenance 
decision-making, a system of decision-making programs covering single bridge to road network level 
has been formed to meet the needs of bridge maintenance at different scales.

There are two main types of innovations in intelligent equipment. (1) during the construction 
process, a large number of sensors are integrated to monitor in real time the external environment 
and structural status of the bridge, such as traffic load, structural stress and other key 
parameters; (2) in the operation and maintenance phase of the bridge, advanced equipment, such as 
unmanned aerial vehicles (UAVs), rope-climbing robots and multifunctional inspection robots are 
used to realize dynamic monitoring of the health status and support management and maintenance 
decision-making.

It can be seen that the development history of bridge management system shows that each significant 
progress is accompanied by innovative breakthroughs in computer technology.

\subsection{Overview of LLM-based Agents Research}
In order to realize AI as a substitute for human labor, LLM-based agents must possess two key 
conditions: first, they should possess advanced comprehension capabilities covering in-depth 
understanding of written language, speech, images, and video, as well as the ability to accurately 
interpret human intentions and generate autonomous feedback. Second, LLM-based agents should have 
the ability to effectively invoke tools or devices, which involves not only direct human manipulation 
of tools, but more advanced LLM-based agents should be able to perceive the physical environment 
through sensors and independently mobilize appropriate tools for real-world self-regulation. This 
section next explores in detail current human exploration and progress in these two areas.

\begin{figure*}[!t]
    \centering{\includegraphics[width=28pc,height=22pc]{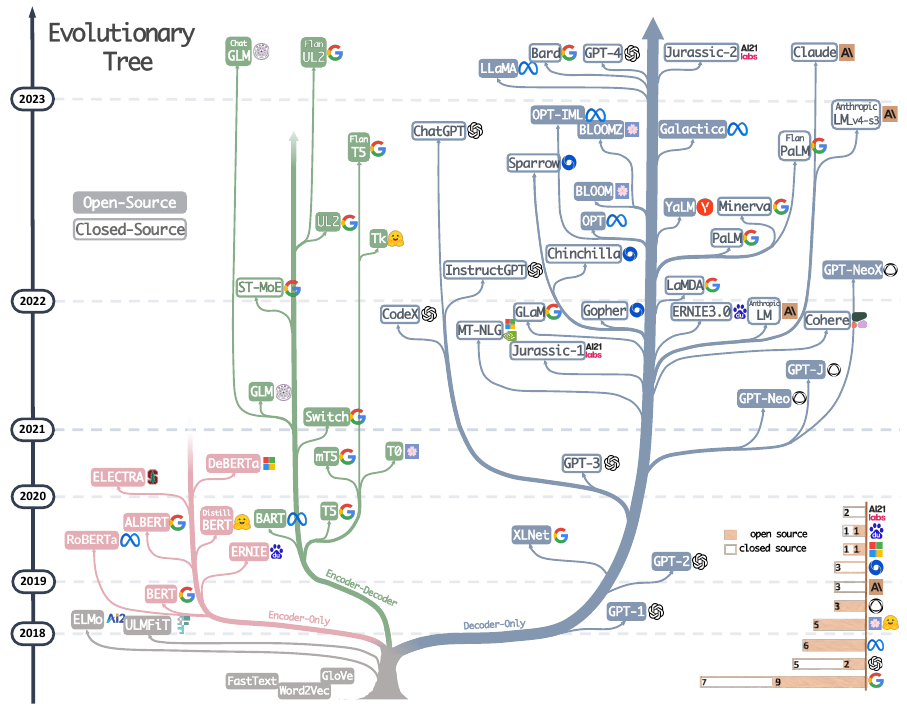}}
    \caption{The evolutionary tree of modern LLMs traces the development of language models in 
    recent years and highlights some of the most well-known models. 
    Summarized by~\cite{yang2024harnessing}. Models on the same branch have 
    closer relationships. Transformer-based models are shown in non-grey colors: decoder-only models 
    in the blue branch, encoder-only models in the pink branch, and encoder-decoder models in the 
    green branch. The vertical position of the models on the timeline represents their release dates. 
    Open-source models are represented by solid squares, while closed-source models are represented 
    by hollow ones. The stacked bar plot in the bottom right corner shows the number of models from 
    various companies and institutions.\label{fig3}}
\end{figure*}

\subsubsection{Larger Language Models}

Language modeling (LM) is the most important task in the field of Natural Language Processing, 
which is essentially a synthesis of Natural Language Understanding and Natural Language 
Generation~\citep{brockopp1983nlp}. 
By enabling computers to learn a large amount of human language data and applying 
specific methods to predict the conditional probabilities between individual tokens (tokens in this 
paper take words in a sentence as an example) in a language, we are able to characterize the logical 
relationships between tokens, thus enabling the language model to understand and generate fluent 
natural language~\citep{nadkarni2011natural}. 
Realizing machines with human-like writing and communication capabilities has been 
a long-term goal pursued by human beings. In academia, the development of language models is usually 
divided into four main stages~\citep{noguer1950key}:

The first stage is the statistical language modeling (SLM) stage. In this phase, the representation 
of words relies on solo heat vectors characterized by a sparse set of vectors representing words
~\citep{turian2010word}. 
Although this approach fails to reflect logical relations in the representation of word vectors, it 
provides an intuitive and understandable way of characterization. As for the characterization of 
sentences, Markov's assumption is used to reveal the logical connections between words in a sentence 
through conditional probabilities. Specifically, this model fixes the context length, which is also 
called n-gram language model, among which the bigram and trigram models are the most common
~\citep{brown1990statistical,kneser1995improved}. This 
phase of language modeling is mainly used for information retrieval
~\citep{lafferty2001document,hiemstra2001using,zhai2008statistical}. However, the main challenge 
faced by statistical language models is the “curse of dimensionality”, i.e., as the vocabulary 
increases, the parameter space required by the model grows exponentially. To alleviate this problem, 
techniques such as smoothing strategies are widely used to mitigate the effects of data sparsity
~\citep{chen1999empirical,rosenfeld2000two}.

The second stage is the Neural Language Model (NLM) stage. Researchers have utilized the word2vec 
method to map words to a low-rank vector space , thus effectively overcoming the sparsity problem 
in the statistical language model~\citep{church2017word2vec,ma2015using,lilleberg2015support}. 
This approach not only enables word vectors to characterize 
logical relations, but also allows intuitive mathematical connections between vectors. Meanwhile, 
through neural network techniques, such as multilayer perceptual machines
~\citep{lilleberg2015support} and recurrent neural networks~\citep{mikolov2010recurrent,
rodriguez1999recurrent,graves2007unconstrained}, 
researchers are able to learn from a large amount of natural language data to 
obtain vector representations of words. This distributed word representation approach has 
demonstrated excellent performance in several downstream tasks of natural language processing (NLP), 
such as machine translation, marking a major breakthrough in the field of NLP in terms of 
representation learning and having a far-reaching impact on subsequent research.

The third stage is the pre-trained model (PLM) stage~\citep{erhan2010does}. 
Unlike the aforementioned Markov chain model, 
PLM utilizes recurrent neural networks to capture word interactions within sentences
~\citep{gers2001lstm,graves2008novel}. 
Subsequently, Google's proposed Transformer architecture introduced self-attention mechanisms and 
positional coding, which greatly improved the model's parallel processing capabilities, enabling it 
to quickly learn a large amount of generalized knowledge and efficiently capture the logical 
relationships between words~\citep{vaswani2017attention}. 
This innovation has given rise to pre-trained language models such as 
BERT and GPT-2. By fine-tuning these pre-trained models on specific tasks, they have achieved 
significant performance gains in almost all downstream tasks of natural language processing (NLP).

The fourth stage is the Large Language Model (LLM) stage. In the current stage, the number of 
parameters of the language model breaks through the level of billions, tens of billions and even 
hundreds of billions. As the model parameters reach new orders of magnitude, the model experiences 
'emergence' phenomenon, and the model performance shows a leapfrog improvement, requiring only a 
small amount of learning to improve the model performance on a specific task
~\citep{beltagy2019scibert}. For example, GPT-3 is 
able to solve specific tasks excellently with simple contextual learning, while GPT-2 performs 
relatively poorly, highlighting the important impact of parameter size on model performance. Along 
with the release of ChatGPT, large-scale language models have become the hottest research in AI, with 
great achievements in areas such as medicine, finance, and autonomous driving
~\citep{wei2022chain}. In 2023, the update 
iteration length of large models reaches the unit of days, and their capabilities in multimodal 
domains such as text, speech, image, and video have been rapidly improved and enriched with features, 
and they have achieved performance beyond the human average in several tasks
~\citep{chang2024survey}.

It can be seen that with the development of the field of Natural Language Processing, especially the 
recent emergence of Large Language Models, computers are equipped with the first condition for the 
realization of LLM-based agents: advanced comprehension capabilities, including in-depth 
understanding of written language, speech, images, and video, as well as the ability to accurately 
interpret human intentions and generate autonomous feedback. This builds a “brain” for LLM-based 
agents, and the next step is to focus on how to equip the brain with the ability to act.

\begin{figure*}[!t]
    \centering{\includegraphics[width=42pc,height=20pc]{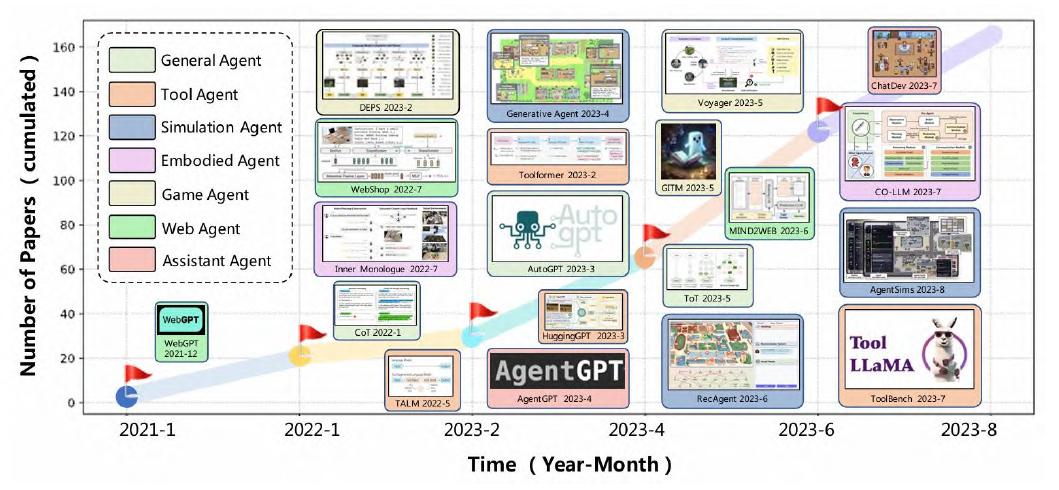}}
    \caption{Illustration of the growth trend in the field of LLM-based autonomous agents.
    \cite{wang2024survey}summarizes the development of intelligibles by time and number of articles. 
    They assign different colors to represent various agent categories. 
    For example, a game agent aims to simulate a game-player, while a tool agent mainly focuses on tool using. 
    For each time period, they provide a curated list of studies with diverse agent categories.
    \label{fig4}}
\end{figure*}

\subsubsection{LLM-based Agents}

Agent belongs to the field of Artificial Intelligence, which aims to construct a human-like 
intelligent entity capable of perceiving the environment in real time, making decisions and taking 
responsive actions~\citep{russell2016artificial}. 
The main difference between LLM-based agents and expert systems is reflected in 
their autonomy, LLM-based agents are able to perceive the environment in real time, rely on the 
knowledge base and chain of thought, generate decisions in line with human expectations, and act 
accordingly~\citep{wooldridge1995intelligent}. 
Humans have always been committed to building LLM-based agents with an intelligence 
level comparable to their own and with the ability to act, and the development of LLM-based agents 
can be divided into four main stages:

The first stage is the symbolic agents stage. In this early stage, researchers worked to create 
entities capable of making decisions~\citep{wilkins2014practical}. 
The initial strategy was to formulate a large number of rules, 
input judgment conditions and feedback into the computer, and make it respond according to these 
rules. During this period, expert systems were representative of symbolic LLM-based agents. The 
advantage of symbolic LLM-based agents is their excellent interpretability, which clearly reveals 
the computer's thought processes~\citep{sacerdoti1975nonlinear,mcallester1991systematic}. 
However, their limitations are equally significant: the limited 
nature of the rule input makes it difficult to handle complex real-world inputs, and as the number 
of rules increases, symbolic LLM-based agents are limited in their speed of response and are unable 
to react quickly to inputs~\citep{kaelbling1987architecture,russell1991right}.

The second stage is the Reactive agents stage. Unlike symbolic agents, LLM-based agents are centered 
on their immediate perception and response to changes in the environment
~\citep{brooks1991intelligence,maes1990designing,nilsson1992toward}. Unlike symbolic LLM-based 
agents, which focus on symbolic manipulation and complex logical reasoning, LLM-based agents focus on 
establishing a direct mapping between inputs and environmental stimuli as well as between outputs and 
behavioral responses~\citep{schoppers1987universal,brooks1986robust}. 
The goal is to achieve accurate and rapid responses with minimal computational resources.

The third stage is the Reinforcement learning-based agents stage. Thanks to increased computational 
power, LLM-based agents are able to handle more complex tasks through interaction with the 
environment~\citep{minsky1961steps}. 
The reinforcement learning framework emphasizes that LLM-based agents must learn to make 
decisions to maximize the cumulative rewards under the constraints of the given tasks and rules while 
interacting with the environment, so as to ensure that the choices made at each step are the optimal 
solutions in the current context~\citep{sutton2018reinforcement}. 
Initially, reinforcement learning LLM-based agents relied on basic 
techniques such as policy search and value function optimization, typically Q-learning
~\citep{watkins1992q} and SARSA~\citep{watkins1989learning}. 
Subsequently, with the advancement of deep learning techniques, deep reinforcement learning has 
emerged, which combines the strengths of both and allows the agents to process high-dimensional 
inputs and to learn higher-level policies, as shown by AlphaGo
~\citep{silver2016mastering} and DQN~\citep{mnih2013playing}. The main advantage of this 
approach is the ability of LLM-based agents to autonomously adapt to unknown environments without 
human intervention. Nevertheless, reinforcement learning still faces challenges of long training 
cycles, inefficient sample utilization, and application in unstable environments.

The fourth stage is the Large Language Model-based agents (LLM-based agents) stage. As mentioned 
earlier, Large Language Models have demonstrated significant performance and can be utilized as the 
brain of LLM-based agents~\citep{kasneci2023chatgpt}. 
LLM-based agents extend perception and action capabilities through 
strategies such as multimodal perception~\citep{zhao2023chat,huang2024language}
 and tool invocation~\citep{kim2023llm,hong2023metagpt}, enhance planning and reasoning 
capabilities through techniques such as thought chaining and problem decomposition
~\citep{hsieh2023distilling,zhang2023multimodal}, and gain the 
ability to interact with the environment by learning from and responding to feedback
~\citep{stamper2024enhancing,huang2023finbert}. LLM-based 
agents acquire generalization capabilities by learning a large training set, thus enabling free 
switching between different tasks, and have been applied to various real-world tasks in areas such 
as financial services, smart homes, education and training, healthcare, and intelligent customer 
service.

It can be seen that LLM-based agents based on Large Language Model can already realize autonomous 
perception, planning, decision-making and action through natural language interaction. Although there 
is still a long way to go before the realization of general artificial intelligence, the current 
research has made computers initially equipped with two conditions for the realization of LLM 
agents-based agents, which not only have advanced comprehension capabilities, but also have the 
ability to effectively invoke tools or devices to perceive the physical environment through sensors 
and independently mobilize the appropriate tools in order to achieve self-regulation of the real world.

\subsection{Why Bridge Operation and Maintenance Need LLM-based Agents}


China has promoted the development of intelligent management of infrastructure through continuous 
technological innovation and engineering practice in the construction and maintenance management of 
highway bridges. However, there are still many problems in the current research: 
\begin{itemize}
    \item In terms of bridge 
    disposition perception, the stability of data collected by sensors is insufficient, and the delay 
    between data collection and data processing of intelligent devices is high, which can not meet the 
    real-time demand
    \item In terms of bridge technology evaluation, there is only a single operation and 
    maintenance scenario description, with a single operation and maintenance disposition indicator and 
    an incomplete evaluation dimension. The accuracy of the existing nature evolution model is low, the 
    degree of intelligence of the performance evaluation method is low, and the false alarm rate of 
    performance warning is high
    \item In terms of bridge operation and maintenance decision-making, the 
    correlation between the data of different dimensions of the bridge operation and maintenance indexes 
    is not strong, and the evolution law of the bridge performance is not clear
    ~\citep{jensen2020innovative}. Disease database is 
    single, and there is a lack of database to support O\&M decision-making. The existing operation and 
    maintenance platform is mainly for data collection, display, equipment control and data management, 
    intelligent, automated decision-making control ability is low, weak interaction ability. 
\end{itemize}

It can be seen that the intelligent level of bridge operation and maintenance is still at a relatively early 
stage, and there are many problems in the state perception, technical evaluation, operation and 
maintenance decision-making, and response to extreme events. The existing operation and maintenance 
platform mainly has basic functions such as data collection, display, equipment control and data 
management, but it is still insufficient in terms of intelligence, automated decision-making control 
and interaction capability. Obviously, the degree of intelligence in the field of bridge operation 
and maintenance is still in the primary stage, and the combination with advanced artificial 
intelligence technology is not close enough.

Therefore, the new generation bridge O\&M management system should incorporate LLM-based agents 
technology to equip each large-span bridge with a personal assistant. At the same time, the system 
should also be based on the road network level, considering the complementary and independent 
relationships between bridge clusters, and realizing the synergy, cooperation, and competitive 
relationships among different LLM-based agents. Studies have shown that coordination among multiple 
intelligences may generate social phenomena among computers, which suggests that the construction of 
single bridge LLM-based agents as well as the synergy and cooperation among bridge cluster 
intelligences at the road network level have full potential to be the future direction of bridge 
operation and maintenance.

\section{Methodology for Constructing Bridge Operation and maintenance LLM-based Agents}\label{sec3}

The integration of Large Language Models and agents is rapidly evolving, but in the O\&M management of 
bridge structures, complex bridge engineering knowledge and specialized O\&M skills are required. The 
acquisition and accurate construction of specialized knowledge is critical for the performance of 
LLM-based agents in downstream tasks. Our goal is to embed specialized knowledge into intelligent 
agents to support automation and LLM-based agents for bridge O\&M management. In this section, we 
describe the methodology for building LLM-based agents ontologies in specialized domains.

\subsection{Knowledge Source}
In order to realize the specialized functions of LLM-based agents, this paper suggests building 
specialized domain intelligences using a combination of distributed knowledge, structured knowledge, 
and multi-round conversation data.

\begin{center}
    \includegraphics[width=0.4\textwidth]{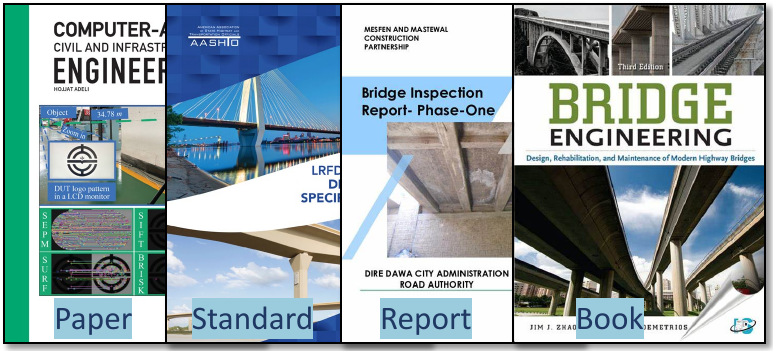}
    \makeatletter
    \def\@captype{figure}
    \makeatother
    \caption{Distributed knowledge can be used to train word embedding models or as external 
    information for retrieval.\label{fig5}}
\end{center}

\subparagraph{Distributed Knowledge}
Distributed knowledge is i.e., natural linguistic expertise.Although current LLM-based agents show 
excellent comprehension and generation capabilities in general-purpose domains, mainstream Large 
Language Models lack specialized domain knowledge in the pre-training dataset, thus making it 
difficult to meet the demands of practical engineering applications. China's Bridge Technical 
Condition Evaluation Standard divides bridge inspections into daily inspections, frequent 
inspections, periodic inspections, and special inspections, and after decades of construction 
in China's bridge engineering field, tens of thousands of inspection reports have been accumulated, 
which can be used as a key source of data for LLM-based agents' specialization. Meanwhile, texts 
such as academic papers, specialized books and industry specifications in the field of bridges are 
also the main sources of data. However, distributed knowledge alone does not 
achieve good results because the complexity and depth of the bridge O\&M domain require LLM-based 
agents to have structured knowledge to address their shortcomings in the depth of knowledge 
representation. Therefore, in the bridge domain, LLM-based agents need to have both high-quality 
distributed and structured knowledge to achieve better performance.

\begin{center}
    \includegraphics[width=0.4\textwidth]{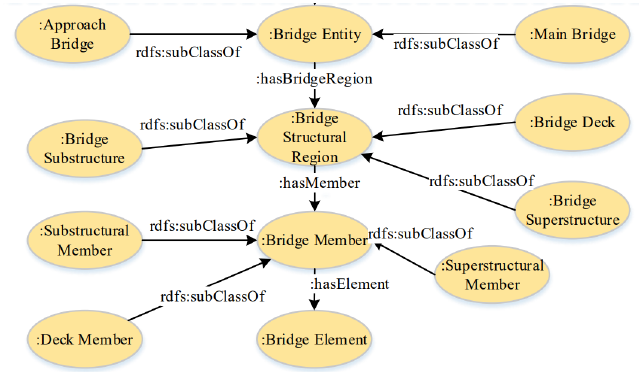}
    \makeatletter
    \def\@captype{figure}
    \makeatother
    \caption{Knowledge graphs as the most typical structured knowledge.\label{fig6}}
\end{center}

\subparagraph{Structured Knowledge}
A typical expression of structured knowledge is the knowledge graph, a concept introduced by 
~\cite{Singhal2012Official}, which is essentially a structured way of organizing knowledge. Knowledge graph can be 
regarded as a kind of knowledge relationship graph, consisting of nodes and edges, where nodes 
represent entities or concepts and edges represent attributes or relationships. The application 
of knowledge graph can effectively solve the problem of the depth of knowledge expression of 
intelligences in the field of bridge O\&M. Distributed knowledge by itself is not sufficient for 
agents to understand the logical relationships in bridge engineering and structural O\&M. It is 
recommended to construct a multilevel O\&M knowledge base covering system-structure-components of 
bridges, perception-evaluation-decision making in O\&M, principle-expression-treatment measures of 
bridge diseases, and preprocessing-threshold setting-abnormal value handling of bridge data in order 
to enhance the accuracy and professionalism of the agents. However, even with the combination of 
distributed and structured knowledge, agents still face challenges in realizing the cyclic dialog 
dataset of inquiry/command-feedback/action.

\subparagraph{Multi-round Dialog Dataset}
Multi-round conversation datasets involve not only linguistic conversations, but alignment between 
LLM-based agents and human commands in a broad sense. At the Large Language Model level, in order to 
understand human intentions, a multi-round Q-A (query-feedback) dataset, i.e., a large amount of 
query and feedback data, is required~\citep{singh2019strings,yi2024survey,abbasiantaeb2024let}. 
In this process, the design of prompt templates (prompt word 
engineering) is crucial, which needs to guide the model to output the desired results without 
updating the weights of the Large Language Model according to the actual needs of the specialized 
domain. At the level of LLM-based agents, in order to realize the automatic maintenance of bridge 
structures, multiple rounds of I-A (instruction-action) datasets, i.e., a large number of instruction 
and action datasets, are required~\citep{esteban2024using}. 
Currently, a variety of intelligent algorithms and advanced 
devices have been developed in the field of bridge operation and maintenance, and the multi-round 
I-A dataset can help LLM-based agents invoke these tools more effectively to accomplish the 
automatic maintenance of bridges. It is worth noting that each tool invocation cannot be supported 
by a large number of algorithms and hardware development techniques.

\subsection{Construction of the Ontology for LLM-based Agents}
Current Large Language Models are generalized Large Models trained by a few companies based on 
Internet data, which perform unsatisfactorily in the professional domain. It is mainly because the 
above Large Language Models lack the training of specialized datasets in the training process. 
Previously, we introduced the sources of training data, and next we introduce the method of building 
LLM-based agents.

\subparagraph{Fine-tuning of LLM}
Fine-tuning of large language models involves further training of pre-trained models using 
domain-specific datasets to optimize their performance on specific tasks
~\citep{otieno2024accuracy,brown2020language}. This process aims to make 
the model better adapted to and perform domain-specific tasks. Fine-tuning involves the tuning of 
both the Embedded Model and the Large Language Model~\citep{ouyang2022training}. 
Fine-tuning of the embedding model involves 
mapping natural language into low-dimensional vectors to express logical relationships between words, 
which requires the use of distributed data to capture specialized noun entities from a large number 
of data sources. Fine-tuning Large Language Models, on the other hand, requires updating model 
parameters with a variety of data, and most open-source models provide official fine-tuning methods 
accordingly. However, fine-tuning Large Language Models requires both strong computational power and 
specialized knowledge, and is accompanied by a certain degree of uncertainty that makes it difficult 
to accurately predict the performance of the model after fine-tuning
~\citep{thirunavukarasu2023large,ding2023parameter,radiya2020fine}. Currently, commonly used 
methods include Retrieval Augmented Generation (RAG) and Chain of Thought (CoT) techniques to guide 
the model to accomplish specific tasks. Although these methods may have limited enhancement of model 
capabilities, they provide better stability.

\subparagraph{RAG and CoT}
Retrieval Enhanced Generation (RAG) is a technique to improve the question and answer quality and 
interaction capabilities of generative AI by utilizing additional data resources without changing the 
parameters of Large Language Models. The workflow of RAG includes loading external documents, 
document segmentation, content vectorization, data retrieval, and finally answer generation
~\citep{lewis2020retrieval}. However, 
a limitation of RAG is that it may lead to retrieval failure of valid results for the case of 
semantically identical but differently worded questions, a problem that decreases as the quality of 
the embedded model improves. CoT significantly improves the performance of Large Language Models by 
guiding them to progressively engage in the process of decomposing a complex question into a series 
of sub-problems and solving them sequentially~\citep{fujita2024llm}. 
A generic template for the CoT technique should 
contain the question, reasoning process, and the answer as three core components to guide the 
reasoning and generation process of the model.

\subparagraph{Langchain build Agent}
LangChain is a state-of-the-art Large Language Model development framework that integrates Large 
Language Models, Embedded Models, Interaction Layer Prompts (Prompts), External Knowledge, and 
External Tools to provide a flexible solution for building LLM-based agents. LangChain consists of 
six core components including Models, Prompts, Indexes, Memory, Chains, and Agents. Indexes, Memory, 
Chains and Agents. These components are connected to each other in the form of chains, enabling 
LLM-based agents to realize autonomous perception, planning, decision-making and action
~\citep{kansal2024langchain}. The 
framework not only empowers LLM-based agents with advanced understanding, but also enhances their 
ability to invoke tools or devices~\citep{m2024augmenting,wang2024survey}. 
LLM-based agents are able to perceive the physical environment 
through sensors and independently mobilize appropriate tools for automation and intelligence in 
bridge O\&M management.

\subsection{Qualitative Result Evaluation}
The previous section of this paper introduced the methods for preparing and constructing the data 
required for LLM-based agents. Based on this, this section will focus on the assessment framework 
and evaluation criteria for LLM-based agents.

Along with the introduction of large-scale language models, several widely adopted evaluation 
standards have been established in the industry, including MMLU~\citep{hendrycks2020measuring}, 
BIG-bench~\citep{srivastava2022beyond}, and HELM~\citep{bommasani2023holistic}, as well as a 
series of human benchmark tests and evaluation tests focusing on model-specific capabilities. In 
addition, there are evaluation benchmarks focusing on model-specific capabilities, such as the 
TyDiQA~\citep{clark2020tydi} 
benchmark focusing on knowledge application and the MGSM
~\citep{qian2022limitations} benchmark focusing on mathematical reasoning. In order to conduct an effective assessment, the selection of benchmarks should be based on the specific objectives of the assessment.

While assessment benchmarks, as previously described, have been widely adopted in evaluating the 
comprehension and production capabilities of large language models, they fail to adequately assess 
the models' ability to plan and make decisions in complex environments, as well as the ability of 
LLM-based agents to act. As LLM-based agents technology advances, there is increasing academic 
interest in assessing the responsiveness of LLM-based agents in complex environments
~\citep{liu2024understanding}.

Currently, while there are no widely accepted benchmarks for the assessment of LLM-based agents, 
researchers have made significant progress in proposing candidate benchmarks that may become future 
assessment standards. For example, ToolBench~\citep{guo2024stabletoolbench} 
is a fine-tuned dataset for evaluating intelligences' invocation of single- and multi-tool commands; 
TE~\citep{aher2023using} evaluates the multifaceted ability of language 
models to simulate human behavior; MetaTool~\citep{huang2023metatool} 
aims to assess whether Large Language Models consciously 
invoke tools and select the appropriate tool for problem solving; and 
LLM-Co~\citep{agashe2023evaluating} focuses on LLM-based 
agents' ability to infer partners' intentions in games, engage in reasoning and the ability to 
cooperate over time. LLM-based agents are usually assessed on a task-specific basis and with a 
certain degree of ambiguity. As LLM-based agents, especially in the field of multi-intelligent 
collaboration, methods for effectively tracking and evaluating the properties of intelligences will 
become increasingly important.

\begin{figure*}[!t]
    \centering{\includegraphics[width=30pc,height=25pc]{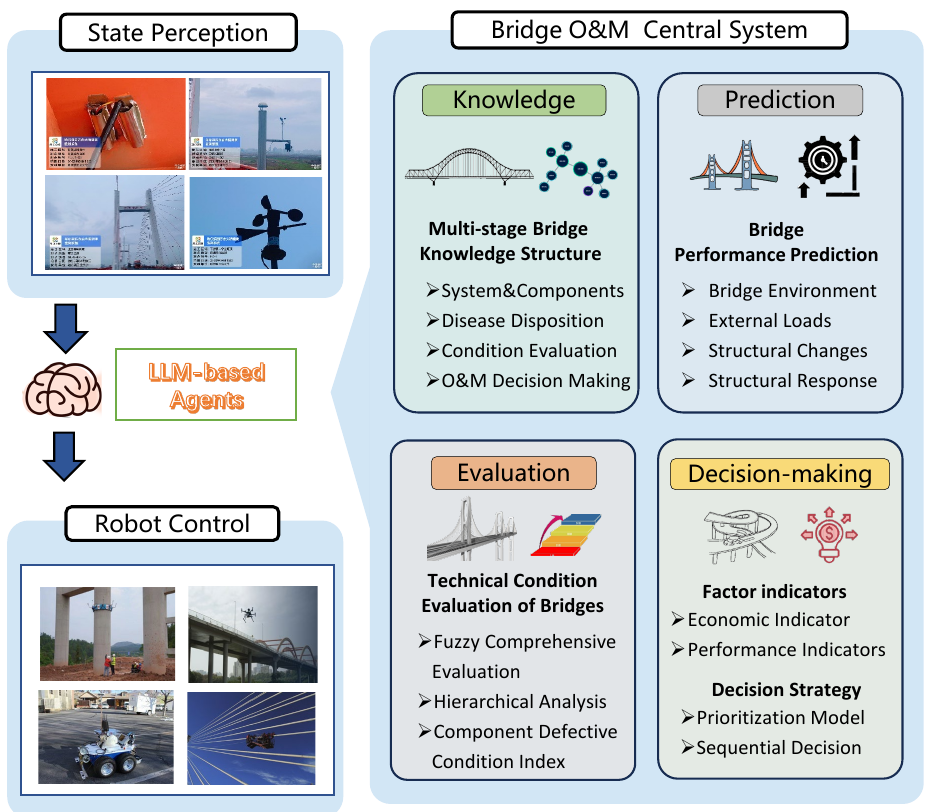}}
    \caption{An example of LLM-based agents framework.The main function of 
    LLM-based agents is to perceive the service state of the bridge, then 
    after the "brain" to think and take action. The "brain" needs to have bridge 
    operation and maintenance knowledge system, can make predictions, evaluation, 
    decision-making on bridge service performance.
    \label{fig7}}
\end{figure*}

\subsection{An Example of LLM-based agent Framework}
This framework aims to build LLM-based agents system that integrates bridge monitoring, operation 
and maintenance perception, indicator prediction, condition evaluation, intelligent decision-making 
and autonomous action. The core of this system lies in the integration of several advanced Large 
Language Models, and the combination of Internet of Things, Artificial Intelligence, Machine 
Learning, and Big Data technologies, to realize the comprehensive monitoring and intelligent 
management of the bridge condition.

\subparagraph{Operations and Maintenance Perception Layer}
(1) Monitoring end equipment. Including various types of sensors (e.g., stress sensors, displacement 
sensors, vibration sensors, environmental monitoring stations, etc.), real-time collection of bridge 
structural health data, environmental parameters (e.g., temperature, humidity, wind speed, etc.) and 
traffic flow information. (2) Data collection and pre-processing. Aggregate the monitoring data to 
the data center through the IoT gateway, and perform pre-processing such as cleaning, compression, 
encryption, etc., to ensure data quality and security.

\subparagraph{Intelligent Processing Layer}
(1) Multimodal data processing. Integrate structured data (e.g., sensor readings) with unstructured 
data (e.g., bridge construction information, maintenance records, historical documents) and extract 
key information using NLP techniques. Understand textual information, such as maintenance reports and 
design documents, based on models such as BERT. (2) Indicator prediction model. Establish a finite 
element model of the bridge, based on physical laws, combined with sequence models such as LSTM, 
Transformer, etc., to predict the damage and remaining life of the bridge and identify potential 
risks. (3) Condition evaluation model. Comprehensive assessment of the bridge's structural 
performance, durability and maintenance needs, evaluating the bridge's service level and functional 
performance, detecting and evaluating the degradation of materials due to environmental factors. 
(4) Intelligent decision-making model. Natural language generation using the GPT series of models to 
develop bridge maintenance, resource allocation, and risk management strategies to assist managers 
in making informed data-based decisions using a decision support system to ensure sustainable bridge 
operation and maximize economic benefits.

\subparagraph{Operations and Maintenance Task Scheduling}
Integrate the model of the intelligent processing layer, construct a knowledge graph, realize 
cross-domain knowledge fusion and complex problem reasoning, transform decisions into specific O\&M 
tasks, and realize the unmanned and autonomous use of structural health monitoring systems, drones, 
robotic cleaning systems, automated inspection vehicles, intelligent monitoring systems, and 
automated maintenance equipment in bridge O\&M, especially in harsh or hazardous environments. 
During the execution of bridge O\&M tasks, the execution effect is continuously monitored and 
feedback data is collected for optimizing subsequent decisions.

\section{Potential Development Directions of LLM-based Agents in Bridge Operation and Maintenance}\label{sec5}
Automation and intelligence of bridge O\&M is the key to improve O\&M efficiency and enhance safety 
and reliability. Through real-time monitoring, intelligent analysis and predictive maintenance, it 
realizes optimal allocation of resources, reduces O\&M costs, and enhances the public's trust in 
bridge safety, which is an important development direction for bridge management in the future. In 
the field of bridge O\&M, the integration of LLM-based agents is expected to reduce labor costs, 
reduce subjective judgments, improve management efficiency, and enable rapid response to emergency 
situations. This section describes some of the research elements that need to be focused on for 
bridge O\&M intelligence.

\subsection{Large Language Model for Bridge Operations and Maintenance Domain}

Currently, Large Language Models in vertical fields are booming. In China, a number of industries 
have launched customized large language models, such as “Hua Tuo” in the medical and health 
industry, “Xuan Yuan” in the financial field, “Han Fei” in the legal industry, and “Red Rabbit” in the field of intelligent customer service and marketing. Through deep learning and natural language processing technologies, these models accurately match 
the needs of various industries, providing a full range of intelligent solutions.

As an important transportation infrastructure, the efficiency of operation and maintenance management 
of bridges is directly related to the safety, smoothness and efficiency of the transportation 
system. The construction of Large Language Model in the field of bridge O\&M and its use as the 
core hub of bridge intelligent management system can realize automated monitoring system and 
intelligent data analysis. The development and application of Bridge O\&M Large Language Model is 
of great theoretical and practical significance to promote the development of bridge O\&M management 
in the direction of intelligence and efficiency.

\begin{center}
    \includegraphics[width=0.4\textwidth]{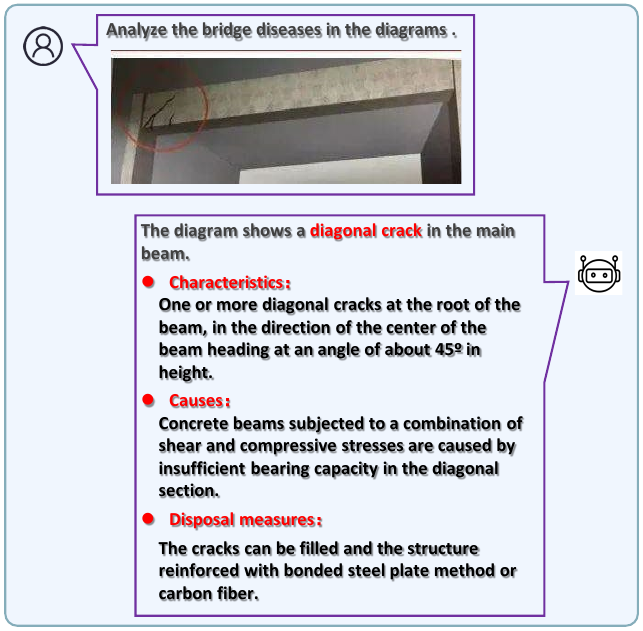}
    \makeatletter
    \def\@captype{figure}
    \makeatother
    \caption{By parsing the pictures, we can analyze the location, main features, causes, impacts 
    and disposal measures of bridge diseases, which cannot be done by the generalized Large Language 
    Models in the market. The recognition of bridge disease pictures by Large Language Model can 
    improve the professionalism and accuracy of bridge operation and maintenance, and reduce labor 
    costs.\label{fig8}}
\end{center}

\subsection{Multimodal Knowledge Graph for Bridge Operations and Maintenance Domain}

Building a multimodal knowledge graph in bridge operation and maintenance is a research and 
development effort aimed at integrating knowledge and data related to bridge operation and 
maintenance, and enhancing the intelligence of bridge management through multimodal information 
processing technology. Large Language Model demonstrates excellent multimodal processing 
capabilities, especially in image and video processing~\citep{koh2024generating,liu2023summary}. 
However, in the field of bridge operation 
and maintenance, the processing capability of existing generalized Large Language Models for images 
is still at a low level. For example, in bridge disease picture recognition, even the best models 
can only recognize bridge cracks with unsatisfactory accuracy. There are a wide variety of bridge 
diseases, including cracked concrete, spalled and exposed reinforcement, weathering, honeycomb, and 
water seepage in reinforced concrete bridges, as well as fracture, corrosion, coating defects, and 
sling relaxation in steel bridges, which cannot be effectively recognized by existing models.

\begin{center}
    \includegraphics[width=0.35\textwidth]{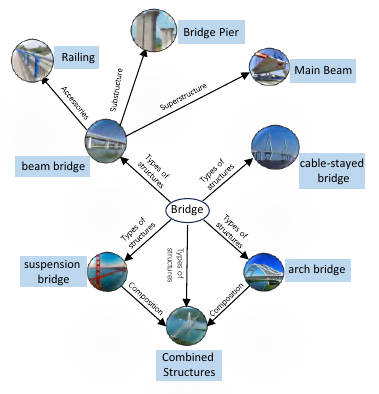}
    \makeatletter
    \def\@captype{figure}
    \makeatother
    \caption{LLM-based agents to realize multimodal (text, picture, video, table) parsing of bridge 
    operation and maintenance information need to establish a knowledge system in various aspects 
    such as bridge structure, disease, and evaluation. Knowledge graph can well increase the 
    understanding depth of the model.\label{fig9}}
\end{center}

To address this problem, combining the multimodal processing capability of Large Language Model with 
the rich data in the field of bridge inspection, constructing a multimodal knowledge graph for bridge 
O\&M, establishing a hierarchical knowledge system, systematically classifying bridge diseases, 
components, parts, and structures, and realizing real-time processing of pictures and videos 
collected on site will be a promising research direction. The construction of multimodal knowledge 
graph in the field of bridge operation and maintenance can not only improve the intelligent level 
of bridge operation and maintenance management, but also provide strong knowledge support for the 
whole life cycle management of bridges.

\subsection{Automatic Generation of Bridge Reports and Assisted Decision-Making}

Currently, most knowledge management Q\&A systems in the field of bridge maintenance are based on 
knowledge graphs~\citep{zareian2020bridging,liu2021knowledge,park2021bridge}. 
Knowledge graph-based Q\&A systems need to extract structured knowledge from bridge 
design, construction, operation, management, inspection, and other data, a process that is 
time-consuming and labor-intensive, and difficult to handle unstructured or semi-structured data
~\citep{chen2020review}. 
In addition, structured knowledge expressions may lead to error transmission and accumulation, and 
the accuracy of knowledge extraction decreases dramatically when data complexity increases.
\\~

\begin{center}
    \includegraphics[width=0.45\textwidth]{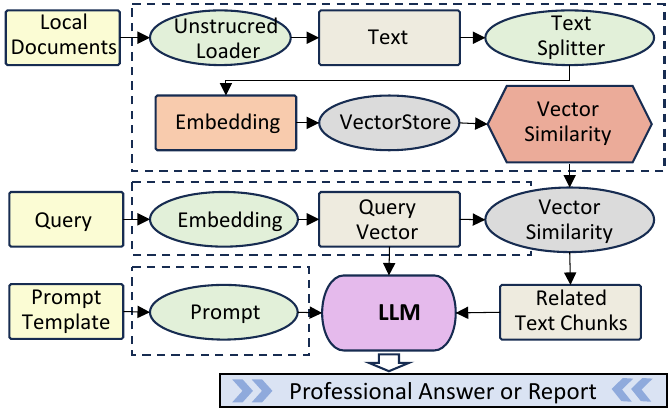}
    \makeatletter
    \def\@captype{figure}
    \makeatother
    \caption{Firstly, data preparation and preprocessing are carried out, and the database data 
    are cleaned, segmented and mapped into word vectors, which are stored in the vector database. 
    When the user question is received, it is preprocessed and mapped into word vectors, and the 
    similarity with the segment vectors in the database is calculated to filter out the relevant 
    information. According to the designed prompt template, the interrogative sentence and related 
    information are input into the Large Language Model to generate a professional answer or report. 
    Finally, user feedback is collected and the process is optimized to improve answer quality and 
    user satisfaction.\label{fig10}}
\end{center}

\subsection{Tools Calling LLM-based Agents}

In the field of bridge operation and maintenance, the introduction of Large Language Model-based 
interaction systems can gradually reshape its traditional paradigm. The Large Language Model-based 
Q\&A system adopts an unsupervised language model learning approach to map the knowledge to a 
continuous numerical space, effectively overcoming a series of problems of traditional knowledge 
extraction methods~\citep{su2019generalizing,shao2023prompting}. 
With its excellent natural language understanding and generation capabilities, 
the system not only greatly improves the intelligence of information retrieval and knowledge 
services, but also promotes the automation and intelligence of operation and maintenance processes. 
O\&M personnel can easily query all kinds of O\&M records and reports through natural language, while 
Large Language Model can quickly provide precise information to assist decision-making, and 
automatically mine potential problems from massive data to enhance the predictability of O\&M. 
Meanwhile, the automated report generation function significantly reduces the time and errors of 
manual report writing and improves work efficiency.

Traditional bridge operation and maintenance management often relies on manual inspection and regular 
testing, which is not only inefficient, but also difficult to comprehensively and accurately grasp 
the actual condition of the bridge. In recent years, with the improvement of bridge inspection 
efficiency and quality needs, the development of intelligent inspection equipment and technology 
has gradually matured, covering drones, rope-climbing robots, underwater robots, sonar detection 
devices, as well as image acquisition technology, laser point-cloud scanning technology, holographic 
photography technology and so on. The future research direction is bound to realize the automation 
and unmanned operation and maintenance management of bridges, and the application of LLM-based agents 
based on Large Language Model is expected to realize this goal.

Firstly, LLM-based agents realize accurate real-time monitoring of bridge structural health and 
environmental parameters by seamlessly integrating and scheduling a variety of monitoring systems 
and sensors, utilizing Large Language Models to quickly analyze the data and discover potential 
anomalies in a timely manner, ensuring the accuracy of monitoring. Secondly, LLM-based agents are 
expected to realize automatic scheduling of advanced equipment such as drones and intelligent robots 
to perform complex tasks such as high-altitude inspection and internal maintenance, realizing 
automation and unmanned maintenance operations. This not only improves maintenance efficiency, but 
also reduces personnel safety risks. These LLM-based agents maintain real-time communication with the 
intelligent body, ensuring accurate execution and efficient management of maintenance work.

\begin{figure*}[!t]
    \centering{\includegraphics[width=38pc,height=19pc]{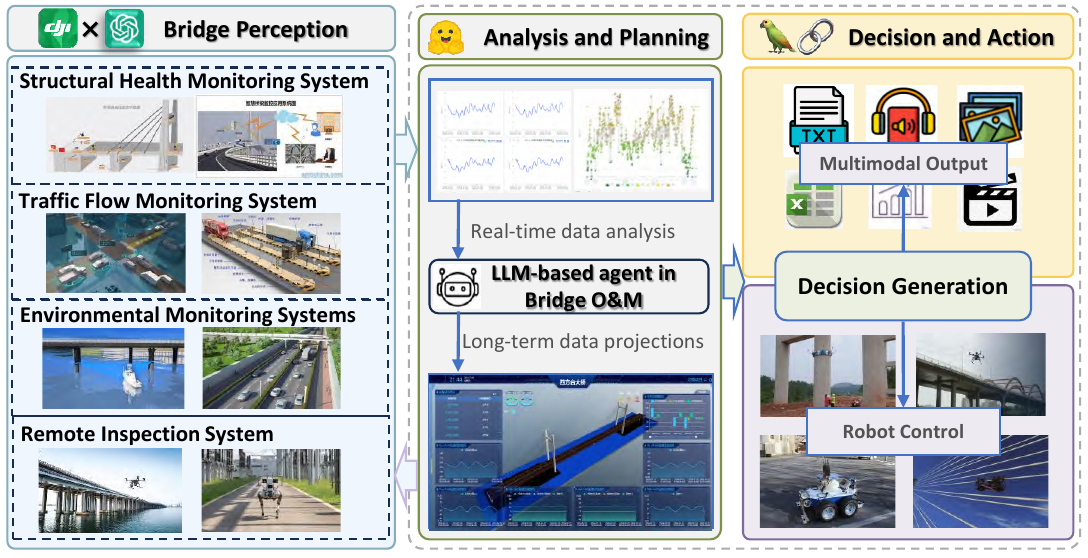}}
    \caption{LLM-based agents first use the monitoring system and remote inspection system to 
    sense the bridge status, then use the multimodal processing capability (text, table, picture, 
    video) to analyze the data in depth through the prediction and evaluation module, and integrate 
    the raw data and its analysis process into the display platform. Eventually, LLM-based agents 
    output decision-making results covering monitoring reports, inspection reports, maintenance 
    programs, and develop action plans to implement bridge operation and maintenance operations by 
    autonomously controlling intelligent devices.The current technology is still difficult to 
    realize the unmanned and automated control of intelligent devices, but this must be the future 
    direction of technological development.
    \label{fig11}}
\end{figure*}

\section{Future Trends and Challenges}\label{sec6}

\subsection{Trend}

\subparagraph{Lightweight and Fast Response}
With the development of large-scale language models, their parameter sizes have reached the billion 
level, showing excellent performance in many aspects~\citep{dong2024multi}. 
However, when applying these Large Language 
Models to bridge operation and management, they face the demands of lightweight deployment and fast 
response. The multimodal monitoring data of bridges are transmitted in real time, and massive data 
are rapidly accumulated, but the computational resources are limited. Therefore, the research focus 
is shifted to the fast identification of heterogeneous multimodal data from multiple sources and the 
development of high-quality, lightweight, and low-latency models, which are crucial to realize the 
application of Large Language Model-based agents in bridge O\&M.

\subparagraph{Flexible tool invocation}
Large Language Model provides LLM-based agents with powerful understanding and reasoning 
capabilities, enabling them to perform autonomous planning and action planning in complex 
environments. However, in the face of diverse intelligent devices in different domains, it is 
necessary to develop unified interfaces to interface LLM agents-based agents, and develop specific 
algorithms for different tools, so as to realize the understanding of semantic information of the 
environment, as well as the recognition of obstacle and target information through the interaction 
between the Large Language Model and the intelligent devices, thus generating appropriate planning 
solutions.

\subsection{Limitations or Challenges}

\subparagraph{Professional synergies and barriers}
The industrialization of AI needs to be closely integrated with professionals in multiple fields, 
and the application of LLM-based agents in the field of bridge operation and maintenance requires 
researchers to have cross-disciplinary knowledge, such as bridge engineering, structural mechanics, 
mechanics of materials, road surveying and geology, as well as the corresponding engineering skills. 
Meanwhile, bridge engineering researchers need to master multi-disciplinary skills such as Natural 
Language Processing, Deep Learning, Machine Learning, Large Language Model, and software and hardware 
development capabilities. This reveals the barriers and challenges in cross-disciplinary 
collaboration.

\subparagraph{Engineering Ethics and Responsibility}
In engineering applications, the health of bridge structures is critical to the transportation 
network and any damage may have social implications. LLM-based agents' recommendations and programs 
may involve loss of property or life, and the issue of responsibility attribution needs to be 
addressed urgently. Currently, LLM-based agents are mainly used as decision aids or action tools. 
In the future, multi-model collaboration will face user privacy and data security challenges, 
requiring multi-party cooperation to establish a sound regulatory and ethical framework.

\section{Conclusion}\label{sec7}
In the course of this study, we have successfully constructed a LLM-based agents, aiming to break 
through the development bottleneck in the field of bridge operation and maintenance and to promote 
the intelligent transformation of this industry. By comprehensively analyzing the current development 
status of the bridge O\&M field and the advancement of LLM-based agents, we conclude that 
LLM-based agents have significant advantages in understanding, generating, planning, 
decision-making, and action, and are able to effectively respond to the challenges facing the bridge 
O\&M field. The framework of LLM-based agents proposed in this study covers several key aspects such 
as data sources, knowledge ontology construction and model evaluation, which provides strong support 
for the intelligent development of the bridge O\&M domain. Meanwhile, we also explore the application 
prospect of this LLM-based agents in the bridge O\&M field, and believe that it has a broad development 
space. However, the application of LLM-based agents to the field of bridge operation and 
maintenance still faces problems such as professional barriers and engineering ethics, which are 
important directions that need to be focused on and solved in future research. Overall, this study 
provides new ideas and methods for the development of intelligence in the field of bridge operation 
and maintenance, which is expected to promote the progress of the whole industry.


\section*{Acknowledgments}

We thank the National Key Research and Development Program of China (2022YFC3801100) for financial support.

\nocite{*}
\bibliography{wileyNJD-APA}%

\end{multicols} 
\end{document}